\documentclass[epj]{svjour}

\usepackage{latexsym}
\usepackage{graphicx}
\usepackage{amsmath}
\usepackage{cite}
\usepackage{amsfonts}
\usepackage{amssymb}

\begin{document}

\title{Adsorption of polyampholytes on charged surfaces}

\author{Florian Ozon\thanks{\emph{Present address: Laboratoire de Dynamique des Fluides Complexes 
(Universit\'e Louis Pasteur - CNRS UMR 7506), 3, rue de l'Universit{\'e}, 67084 Strasbourg Cedex, France}} 
\and Jean-Marc di Meglio\thanks{\emph{Universit\'e Louis Pasteur and Institut Universitaire de France}}
\and Jean-Fran\c cois Joanny\thanks{\emph{Present address: Physico-Chimie Curie (Institut Curie - CNRS UMR 168), 
11 rue Pierre et Marie Curie, 75231 Paris Cedex 05, France}}
}                 

\institute{Institut Charles Sadron (CNRS UPR 022), 6 rue Boussingault, 67083 Strasbourg Cedex, France}
\date{Received: date / Revised version: date}
%
\abstract{We have studied the adsorption of neutral polyampholytes on model charged surfaces 
that have been characterized by contact angle and streaming current measurements. The loop size distributions of adsorbed polymer chains have been
obtained using atomic force microscopy (AFM) and compared to recent theoretical predictions. We find a qualitative agreement with theory;
the higher the surface charge, the smaller the number of monomers in the
adsorbed layer, in agreement with theory.  We propose an original scenario for the adsorption of polyampholytes 
on surfaces covered with both neutral long-chain and charged short-chain thiols.
} 
\maketitle

\section{Introduction}
\label{sec:intro}

The study of the adsorption
of charged polymers on charged surfaces is of great
interest in both polymer physics and industrial practice. A better
understanding of the adsorption phenomenon has already led to numerous
applications related to wetting, lubrication, adhesion\ldots  Polyelectrolytes are for instance currently used to control the colloidal stability of
dispersions or in waste water treatment. Relatively few experimental studies have however been performed on polyampholyte adsorption. Polyampholytes
are polymers carrying both positive and negative
charges along the same chain \cite{blaakmeer, peffer, kato, leberre}. On the theoretical side, Dobrynin,
Rubinstein and Joanny have recently proposed a new adsorption mechanism induced by the polarization of the chains in the electric field of the surface \cite{dobrynin}.

The present work aims at comparing theory and experiments by characterizing
adsorbed polyampholyte layers on controlled charged surfaces with atomic force microscopy (AFM).
Section \ref{sec:exp} presents the experimental procedures: the fabrication of the charged surfaces by reaction
of neutral and charged thiols on gold-coated substrates, the AFM experiments and the two characterization techniques of
the surfaces, contact angle and streaming current measurements. Section \ref{sec:theory} summarizes
the theory of polyampholyte adsorption: the case of a single chain \cite{dobrynin} has been extended to the case of a
solution \cite{obukhov} and the monomer concentration profile allows us to derive
the theoretical loop size distribution in the adsorbed layer. In Section \ref{sec:res},
the experimental loop size distribution is compared to theory and qualitative agreement is found. Contact angle and electrokinetics
measurements are interpreted in the light of the AFM results.

\section{Experimentals}
\label{sec:exp}

\subsection{Surface preparation}
The charged surfaces have been prepared as described in \cite{bain1}.
Substrates were cleaved mica for AFM experiments and contact angle
measurements, or microscope glass slides for streaming current measurements. The
gold-coated surfaces have been realized under high vacuum by deposition of 5 nm of
chromium, which acts as a wetting agent, followed by 50 nm of gold. The surfaces
are then left about 20 hours in an ethanol millimolar solution of thiols.
We have used {HS-(CH$_2$)$_2$-SO$_3^-$ Na$^+$, HS-(CH$_2$)$_2$-CH$_3$}
(both purchased from Aldrich) and {HS-(CH$_2$)$_2$-(CF$_2$)$_7$-CF$_3$}
(kindly provided by Marie-Pierre Kraft from ICS-Strasbourg). The surface charge is changed by mixing charg\-ed and 
neutral thiols in varying proportions. Table \ref{thioltabl} gives
their respective lengths when the aliphatic (or fluorinated) chain is
extended.
\vspace{0.5cm}

\begin{table}[ht]
\begin{center}
\begin{tabular}[c]{|c|c|}
\hline
Thiol & Size (nm)\\
\hline
HS-(CH$_2$)$_2$-SO$_3^-$ & 0.6\\
\hline
HS-(CH$_2$)$_2$-CH$_3$ & 0.6\\
\hline
HS-(CH$_2$)$_2$-(CF$_2$)$_7$-CF$_3$ & 1.5\\
\hline
\end{tabular}
\vskip 5mm
\caption{Lengths of thiols in all-trans conformations}
\label{thioltabl}
\end{center}
\end{table}

\subsection{Polyampholyte adsorption}
\begin{figure}[ht]
\centering{\resizebox{0.5\textwidth}{!}{
  \includegraphics{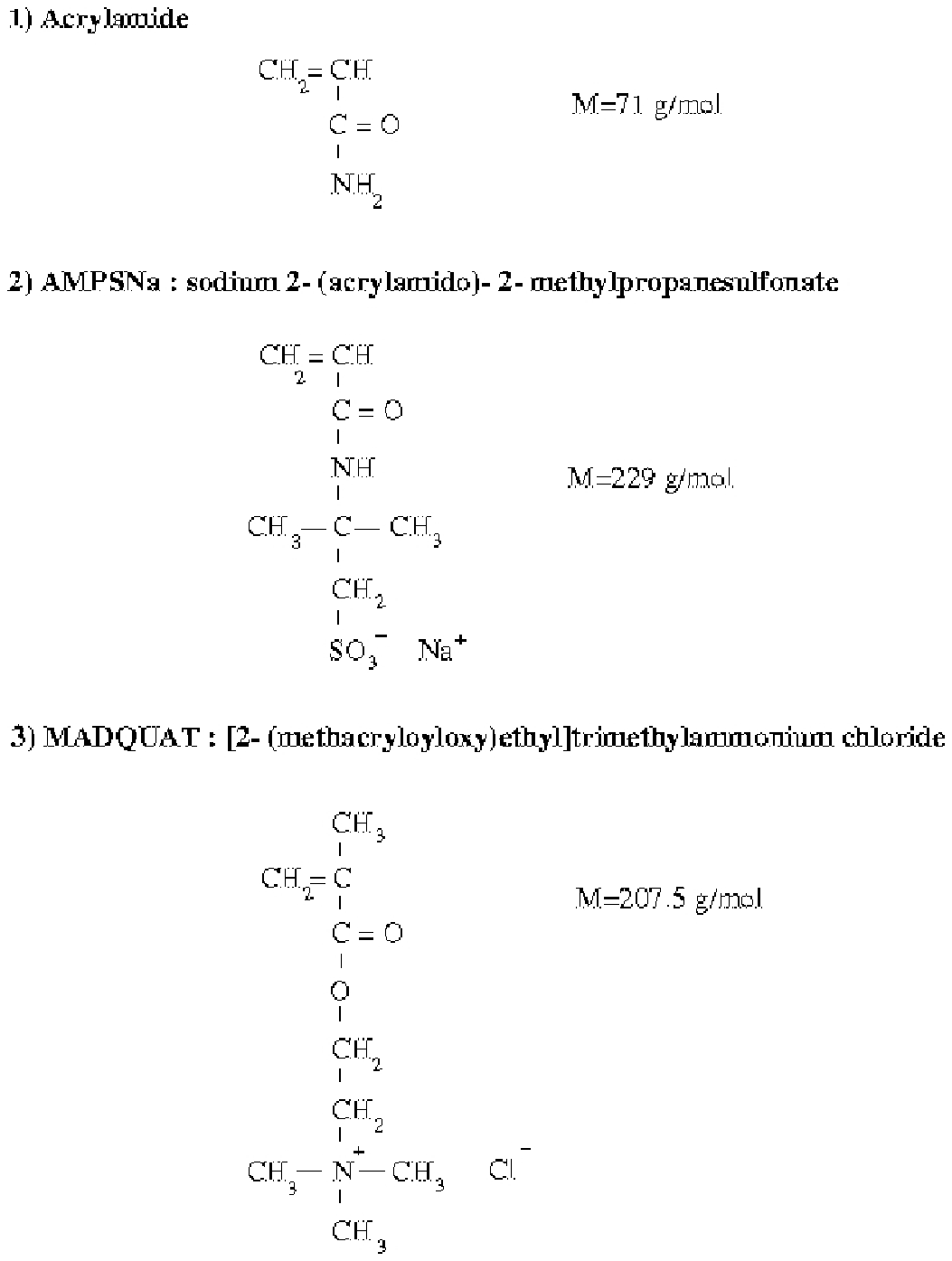}
}}
\caption{Chemical formulae of the monomers used for polyampholyte synthesis.}
\label{monomers}
\end{figure}

The polyampholyte has been synthesized
by  microemulsion-polymerization  by S. Neyret \cite{microemul} from a neutral monomer,
acrylamide, and two charged monomers, sodium
2-(acryl\-amido)-2-me\-thyl\-propane\-sulfonate (AMPSNa) and
[2-(me\-tha\-cryl\-oxy)\-ethyl]tri\-me\-thyl\-ammo\-nium chlor\-ide (MAD\-QUAT). The
chemical formulae are shown on Figure \ref{monomers}.
The total number of monomers in the polymer is $\simeq 4\,10^{4}$; its
composition (93.5/3.75/3.75 \% in
acrylamide/AMPSNa/MAD\-QUAT) has been determined by elementary chemical
analysis and shows that the polyampholyte is globally electrically neutral.
The experimental determination of reactivity ratios has shown that polymerization in microemulsion
leads to an almost random distribution of monomers along the chain \cite{microemul}.
This is due to the strong concentration of counterions in the microemulsion droplets
(diameter $\simeq$ 50-100 nm) that strongly screens the electrostatic interactions between
charged mono\-mers. However, because of the different reactivity ratios of the monomers, the
fluctuations of the net charge of the polyampholytes (defined as the sum of all the charges
along the chain), are larger than  predicted for a
perfectly random distribution \cite{diffusion}. After the synthesis, the
counterions are removed by dialysis against
pure water. The polyampholyte solution has been
prepared by dissolving 10 mg of polymer in 100 g of a  NaCl aqueous solution (concentration 0.15 mol.l$^{-1}$), the polymer being insoluble in pure water and
the critical salt concentration to reach solubility, as determined in \cite{neyret},
being 0.1 mol.l$^{-1}$. The thiolated surfaces have been left in the polyampholyte
solution for one hour, this was long enough for the polymer to adsorb. After that, the surfaces have been rinsed with milli-Q
water before being used in the AFM experiments.  
While salt is needed in bulk solution to ensure that polyampholyte chains do not collapse, the electric field of a charged substrate should be sufficient to stretch adsorbed chains ({\em cf} Fig. \ref{config}) and avoid their collapse; loops of adsorbed polyampholyte can then develop in milli-Q (de-ionized) water. This has been addressed theoretically in \cite{netz} where it is suggested that the behaviour of an adsorbed polyampholyte is similar to the behaviour of an adsorbed soluble polymer.
Throughout our study, we consider water as a $\theta$-solvent for the polyampholyte;
water is indeed close to $\theta$-solvent for polyacrylamide (the second virial coefficient $A_{2}$ for a polyacrylamide with a molar mass equal to
$4.7\,10^6$ g.mol$^{-1}$ is low: $A_{2}\simeq0.64\,10^{4}$ mol.cm$^{3}$.g$^{-2}$ \cite{viriel}) and the used polyampholyte is mostly composed of
acrylamide.
\subsection{AFM experiments}
A Digital Instruments Nanoscope III force
microscope has been used, fitted  with its standard fluid cell. For each
experiment, a new cantilever was cleaned in a water plasma \cite{plasma}. Cantilever spring
constants (from Digital Instruments specifications) were 0.05 N.m$^{-1}$ for experiments on {HS-(CH$_2$)$_2$-SO$_3^-$/\- HS-(CH$_2$)$_2$-CH$_3$} surfaces and 0.03
N.m$^{-1}$ for experiments on {HS-(CH$_2$)$_2$-SO$_3^-$/ HS-(CH$_2$)$_2$-(CF$_2$)$_7$-CF$_3$} surfaces. Milli-Q
water (pH~=~5.7) is injected in the liquid cell set-up and we have then waited half an hour to start the AFM force measurements. The tip-surface then underwent repeated approach-separation cycles
at 1 Hz frequency; beyond this value, hydrodynamic forces appear. In
de-ionized water (pH = 5.7), the total salt concentration (H$_3$O$^+$ and
HCO$_3^-$) is $4\,10^{-6}$ mol.l$^{-1}$, which corresponds to a Debye
length $\kappa^{-1}\simeq150$ nm. The thickness of the adsorbed layer predicted by theory being roughly
$\kappa^{-1}$ in the pseudo-brush regime, a $z$-displacement of the
piezoelectric system much larger than this length was chosen, namely 500 nm, which allows to determine precisely the force profile
over a 400 nm range.

\begin{figure}[ht]
\centering{\resizebox{.45\textwidth}{!}{
  \includegraphics{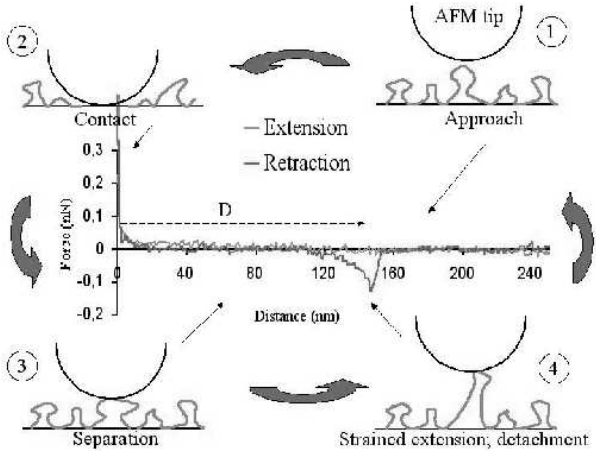}
}}
\caption{Schemes showing a possible path to the formation and extension of a connective bridge (after \cite{senden}); corresponding force curve.}
\label{cycle}
\end{figure}

A peak sometimes appears on the force curves and is understood as the
detachment of a loop that bridged tip and  substrate \cite{senden,silber}(Figure \ref{cycle}). Some extension
curves have been fitted using the Freely Jointed Chain \cite{fjc} and the
Worm Like Chain \cite{wlc} models and the extension of the
loop at the detachment was always found above 90 \% of the contour length; the detachment distance
of the loop can  be considered as its contour length.
 
For each AFM experiment, more than two hundred jumps have been collected to obtain the loop size distribution. The data have been analyzed using the
method introduced in \cite{senden} for the adsorption of polydimethylsiloxane on silica, 
and later used to investigate the adsorption layers of polyacrylamide derivatives \cite{silber}. The loop size distribution $S(n)$ (defined as the
probability for a loop to have $n$ monomers) is deduced from the distribution $p(D)$ of
detachment distances $D$  via the relation $S(n)\,dn = p(D)\,dD$ and the hypothesis that $D\simeq na$ where $a$ is the Kuhn length.
Assuming that the experimental situation corresponds to the {\em fence} regime defined in Section \ref{sec:theory} and using Eq. (\ref{Sn}),
the distribution $p(D)$ is fitted by:
\begin{equation}
p(D) = A\,D^{-1/2}(D+h)^{-1}\label{eqfit}
\end{equation}
$A$ is a numerical prefactor and the number $g$ of monomers in  a chain section with a size 
of the order of the layer of thickness $\lambda$ (the Gouy-Chapman length, see Section \ref{sec:theory}) is given by
$h/a$. Values of $D$ lower than 20 nm have not been taken into account
in the fit because of a lower resolution near the surface and the presence of
a primary adhesion peak (due to van der Waals interactions) on some force curves.

\subsection{Contact angle measurements}
Contact angles of a water sessile
drop have been measured using an optical microscope. The surfaces were tilted until
the drop moved, defining  both advancing $\theta_{a}$ and receding 
$\theta_{r}$ angles. Because of the Young-Dupr\'{e} relationship $\gamma_{LV}\cos\theta=\gamma_{SV}-\gamma_{SL}$, \ where $\gamma_{LV}$, $\gamma_{SV}$ and
$\gamma_{SL}$ are respectively the interfacial free energies of the
liquid-vapour, solid-vapour and solid-liquid interfaces, $\cos\theta$ was
plotted against the ratio of charged thiols to (charged + uncharged) thiols. Indeed, for a given liquid/vapour couple (here water/air), $\cos\theta$ should approximately vary linearly
with the solid surface composition if there is neither segregation of the thiols at the surfaces nor special conformational effects of the thiols. The
contact angle hysteresis $(\cos\theta_{r}-\cos\theta_{a})$ also gives some insight on the surface  heterogeneities \cite{hyster1,hyster2, hyster3}.

\subsection{Electrokinetic measurements}

\begin{figure}[ptb]
\centering{\resizebox{.45\textwidth}{!}{
  \includegraphics{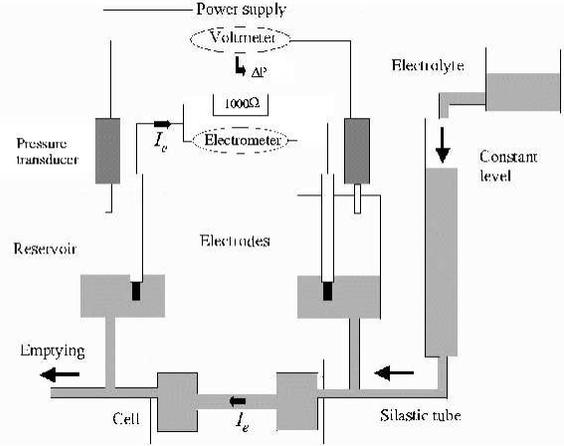}
}}
\caption{Schematic set-up for streaming current measurements.}
\label{apparatus}
\end{figure}

\begin{figure}[ht]
\centering{\resizebox{0.45\textwidth}{!}{
  \includegraphics{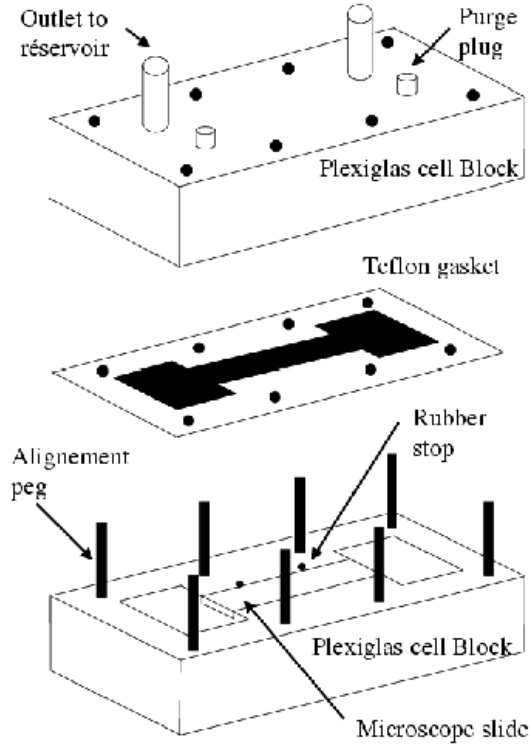}
}}
\caption{Detailed view of the measurement cell.}
\label{cell}
\end{figure}

An original apparatus to measure the streaming current of planar substrates has been designed and is schematically represented on Figure \ref{apparatus}. Figure \ref{cell} shows a detailed view of the measurement cell.

The measurement cell is built from two PMMA plastic blocks that comprise an
accurately cut depression to fit in two test glass slides of dimensions 7.5
x 2.5 x 0.1 cm$^3$. Each slide is firmly held in position by two Viton
rubber stops. A small reservoir was machined at each extremity of the base block.
The two blocks are separated by a Teflon (PTFE) gasket of 0.02 cm thickness,
aligned with stainless steel pegs.
A capillary of dimension 7.5 x 2.2
x 0.02 cm$^3$ is then formed between the two slides by clamping together the two blocks
with 8 screws.
The principle of the measurement is the
following: upstream, an electrolyte solution ($10^{-3}$ mol.l$^{-1}$ KCl at pH = 5.7)
falls into a column so that the level of the liquid remains constant, which
ensures a constant pressure difference $\Delta p$ between the two extremities of the
measurement cell; $\Delta p$ is measured via two Kobold pressure transducers (0-$10^4$ Pa) linked to tightly closed reservoirs. The
measured pressure difference accuracy is 5 Pa. The streaming current is measured with a Keithley 617 electrometer (impedance
$2\,10^{14}$ M$\Omega$) used as a nano-amp\-ere\-meter connected to two silver
electrodes placed in each reservoir of the measurement cell. The electrodes were covered with
silver chloride by electrolysis to prevent their polarization. 

The principle of the $\zeta$ potential determination has been discussed by Hurd and Hackerman
\cite{hurd} in their study of metals (Au, Ag and Pt) in
aqueous solutions. In order to measure the streaming current, we have used a shunt resistance of $10^{3}$ $ \Omega$; the measured impedance of the
capillary tube is 1 M$\Omega$ which ensured that the measured current did not depend on the value of the shunt resistance \cite{hurd}.

In the case of a laminar and established flow across a
charged capillary of length $L$, width $l$ and height $h$, the $\zeta$
potential can be calculated from the pressure $p$ dependence on the streaming
potential $U_s$ and the streaming current $I_s$ (Smoluchowski equations
\cite{hunter}):
\begin{equation}
\zeta(U_{s})=\frac{\eta\left[  \lambda_{l}+\dfrac{2\lambda_{s}}{h}\right]}{\varepsilon}\frac{dU_{s}}{d(\Delta p)}
\label{zetaU}
\end{equation}

\begin{equation}
\zeta(I_{s})=\frac{\eta}{\varepsilon}\frac{L}{lh}\frac{dI_{s}}{d(\Delta p)}
\label{zetaI}
\end{equation}

\noindent where $\zeta(U_{s})$ is the $\zeta$ potential calculated from the streaming
potential, $\zeta(I_{s})$ the $\zeta$ potential obtained from the streaming
current, $\eta$ the viscosity, $\varepsilon$ the permittivity of the
fluid, $\lambda_{l}$ the liquid (electrolyte) conductivity and $\lambda_{s}$
the surface conductivity. If the influence of the interfacial conductivity is
not considered, only an apparent $\zeta$ potential is determined from the
streaming potential which is lower than the actual one. The $\zeta$ potential
determined from the streaming current, however, is independent of the surface
conductivity and is a function only of the channel geometry for a given
electrolyte. The thiolated gold surfaces are {\it a priori} conducting and we have then chosen to deduce
$\zeta$ from streaming current measurements.

\section{Theory}
\label{sec:theory}
We summarize in this section the theory of polyampholyte adsorption in a $\theta$-solvent \cite{dobrynin,obukhov} that we need to derive the loop size
distributions. We also give numerical estimates of the various threshold values using the characteristics of the polyampholyte chains
used in our experimental study and described in Section \ref{sec:exp} ($N$, number of monomers $\simeq 4\,10^4$; $f$, fraction of charged monomers $\simeq0.075$; $a$, monomer size $\simeq 0.3$ nm; $R_0$, ideal (Gaussian) chain size $\simeq60$ nm).

\subsection{Single chain adsorption \cite{dobrynin}}
In the Poisson-Boltzmann approximation, a charged surface (of surface charge density $\sigma_0$), immersed in a solvent of dielectric permittivity 
$\varepsilon$, in absence of added salt,  
generates an electric field $E(z)$ at a distance $z$ from the surface:
\begin{equation}
E=\frac{e\sigma_0}{\varepsilon(1+z/\lambda)}
\end{equation}
where $\lambda$ is the so-called Gouy-Chapman length which gives the thickness of the counterion layer close to the surface \cite{israelachvili}:
\begin{equation}
\lambda=\frac{1}{2\pi\sigma_0 l_{B}}
\end{equation} 
with $l_B$ the Bjerrum length $l_B= e^2/(4\pi \varepsilon k_T)$  ($e$ is the elementary charge and $k_BT$ the thermal energy, $l_B \simeq 7$~\AA~in pure water).

Because of the statistical distribution of the charges along the chain,
the two halves of a neutral chain carry opposite charges of order $(fN)^{1/2}e$.
The electric field can polarize a {\em neutral} polyampholyte chain which behaves as a dipole (of size $R_z$ and charge $e(fN)^{1/2}$)
 and is then attracted by the charged surface. In a $\theta$-solvent of the polymer backbone, the chain is stretched when $R > N^{1/2}a$;
this corresponds to a threshold field  $E_{1}\simeq 1/(N^{1/2}a)\,(k_BT/e)\,\,(fN)^{-1/2}$.

\begin{figure}[ht]
\centering{\resizebox{0.45\textwidth}{!}{
  \includegraphics{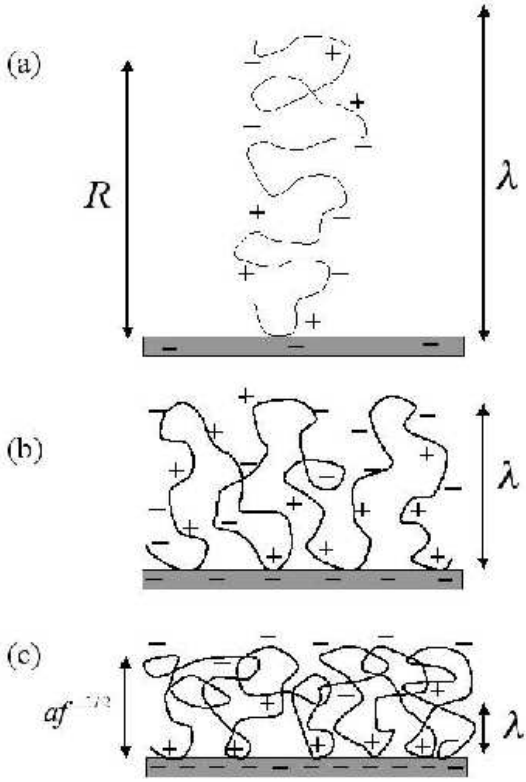}
}}
\caption{Schematic sketch of the configurations of a polyampholyte chain near a
charged surface: (a) the pole regime; (b) the fence regime; (c) the pancake
regime.}
\label{config}
\end{figure}

Three regimes of adsorption have been predicted depending on the charge density $\sigma_0$ (Figure \ref{config}).

i) pole regime ($\lambda\gg aN^{1/2}$). 
A chain is stretched when $E>E_{1}$; the corresponding surface charge density and Gouy-Chapman length are 
respectively $\sigma_{1}\simeq (fN)^{-1/2} (l_{B }R_{0})^{-1}$ and $\lambda_{1}\simeq R_{0}(fN)^{1/2}.$ For the
polyampholyte studied in the experiment, we get $\sigma_{1}\simeq6.9\,10^{13}$ charges/m$^{2}$ and $\lambda_{1}\simeq 3290$ nm.
Beyond this threshold, the chain size $R_{z}$ in the direction normal to the surface
grows linearly with charge density as $R_{0}\sigma_0/\sigma_{1}.$

ii) fence regime.
The Gouy-Chapman length $\lambda$ becomes comparable with $R_{z}$ when $\sigma_0\simeq \sigma_{2}\simeq (fN)^{-1/4}(l_{B}R_{0})^{-1}$.
Because of the screening of the electric field by the counterions, 
the chain then remains confined within a slice of height $\lambda$ (for our system, $\sigma_{2}\simeq5.2\,10^{14}$ charges/m$^{2}$
associated with a Gouy-Chapman length $\lambda_{2}\simeq440$ nm). An adsorbed chain can be divided into subunits (blobs) of size $\lambda$.
The number $g$ of monomers per blob is obtained by considering each subunit as an independent chain:
\begin{equation}
g\sim f^{-1/3}(l_{B}a\sigma_0)^{-4/3}
\label{g}
\end{equation}
Each blob carries an induced dipole moment $e\lambda f^{1/2}g^{1/2}$.

iii) pancake regime.
The polyampholyte chain is strong\-ly bound to the surface when the adsorption energy of each blob of size
$\lambda$ is larger than the thermal energy $k_BT$. This happens for a charge density 
$\sigma_0$ larger than $\sigma_{3}\simeq f^{1/2}/(al_{B})$. The condition $\sigma_0>\sigma_{3}$ then defines the pancake regime:
the Gouy-Chapman length $\lambda$ is smaller than the
mean-square distance between charged monomers $af^{-1/2}=\lambda_{3}$ ($\sigma_{3}\simeq2.3\,10^{17}$ charges/m$^{2}$ and
$\lambda_{3}\simeq1$ nm in our experimental situation). Monomers with charge opposite to
the surface charge are in contact with the surface while the other monomers form loops dangling in
solution at distances $z>\lambda$. The average size of these loops can be estimated from the balance between
the electrostatic force acting on the monomers $\simeq e^{2} \sigma_0\lambda/(\varepsilon z)\simeq k_BT/z$ and the 
elastic force $\simeq k_BTzf/a^{2}$ needed  to stretch
a chain with $f^{-1}$ monomers; the thickness of the adsorbed layer is then approximately $af^{-1/2}$. It is interesting to
note that this thickness does not depend on the surface charge $\sigma_0$.

\subsection{Multichain adsorption \cite{obukhov}}
\begin{figure}[ht]
\centering{\resizebox{.5\textwidth}{!}{
  \includegraphics{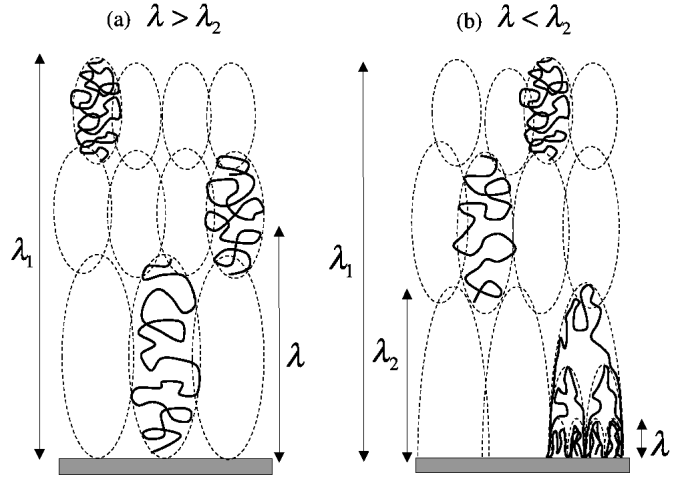}
}}
\caption{Schematic sketch of the configurations of polyampholyte chains in the adsorbed layers.\ a) Multilayer of stretched chains. (b) Self-similar stretched 
pseudo-brush at distances from the surface $z<\lambda_{2}$ and multilayer of stretched chains at $\lambda_{2}<z<\lambda_{1}$.}
\label{psbrush}
\end{figure}

The previous model has been extended to the case of
multichain adsorption by Dobrynin and co-workers \cite{obukhov}. 

i) multilayers of stretched chains ($\sigma_1<\sigma_0<\sigma_{2}$, Figure \ref{psbrush}.a).

The predicted monomer concentration profile is:
\begin{equation}
c(z)\simeq c_{0}^{\ast}\frac{\lambda_{1}}{\lambda+z}
\label{profilepole}
\end{equation}
where $c_{0}^{\ast}= a^{-3}N^{-1/2}$ is the overlap concentration. At
distances $z>\lambda$, a hyperbolic
density profile $c(z)\sim c_{0}^{\ast}\lambda_{1}/z$ is expected. Near the
surface, the polymer density saturates at $c(0)\simeq c_{0}^{\ast}\lambda_{1}/\lambda$. 
The crossover between semi-dilute and dilute regimes
of the adsorbed chains occurs at distance $z\simeq\lambda_{1}$ where the attractive interaction between
each chain and the surface compares to $k_BT$: $\lambda_{1}$ can be considered as the
thickness of the adsorbed layer.

ii) self-similar stretched pseudo-brush ($\sigma_2<\sigma_0<\sigma_{3}$, Figure \ref{psbrush}.b). 

The size of the
chains in the first adsorbed layer saturates at $\lambda_{2}\simeq R_{0}(fN)^{1/4}$. 
The adsorbed layer at distances
$z<\lambda_{2}$ can be viewed as a self-similar pseudo-brush of stretched
polydisperse loops. 
The stretching of a blob of size $z$ containing $g(z)$ monomers can be estimated by balancing the
elastic energy  $k_BTz^{2}/(a^{2}g(z))$ with the polarization
energy $e\sqrt{fg(z)}zE(z)$. For $z>\lambda$ the electric field is
$E(z)\simeq e/(\varepsilon l_{B}z)$. The equilibrium density profile is given by:
\begin{equation}
c(z)\sim a^{-3}\left(  \frac{af}{z+\lambda}\right)^{1/3}
\label{profilefence}
\end{equation}

For $z>\lambda_{2}$, the concentration profile is still given by equation (\ref{profilepole}).
The total thickness of the adsorbed layer is $\lambda_{1}$.

iii) $\sigma_0>\sigma_{3}$. The
size of the polyampholyte molecules in the first layer is $\lambda_{2}$,
but these chains are self-similarly stretched at all length scales between $\lambda_{3}$ and $\lambda_{2}$. 
In the layer of thickness
$\lambda_{3}$ near the wall the monomer concentration 
is constant and proportional to $a^{-3}f^{1/2}$. At distances between
$\lambda_{2}$ and $\lambda_{1}$ one finds a multibrush of stretched chains.

\subsection{Loop size distribution inside the pseudo-brush for $\sigma_2<\sigma_0<\sigma_3$}
\label{sec:analyse}
We derive in this section the loop size distribution in the regime of interest for our experiment.
The same treatment can be used to obtain the loop size distribution in the other regimes. 

We consider a layer of thickness $dz$ located at a distance $z$ from the charged surface; the number
of monomers (per unit area) in this layer is \cite{degennes}:
\begin{equation}
c(z)\,dz=p(n)\,dn\label{relationczpn}
\end{equation}
where $c(z)$ is the monomer concentration and $p(n)$ the probability (per unit
area) that a monomer belongs to a loop comprising more than $n$ monomers. The loop
size distribution $S(n)$, probability to find a loop with n monomers, is then derived from $p(n)$ by:
\begin{equation}
S(n)=-\frac{\partial p}{\partial n}\label{loopdistri}
\end{equation}
The relation between $z$ and $n$ is found by balancing the elastic energy of a
strand of size $z$ containing $n$ monomers and its polarization energy in the
electric field $E(z)$:
\begin{equation}
\frac{k_BT}{na^2}z^2 \simeq\frac{e\sigma_0}{\varepsilon}\,\frac{\lambda}{z+\lambda}\,e(fn)^{1/2}\,z
\end{equation}
leading to:
\begin{equation}
z(z+\lambda)\sim n^{3/2}\label{relationzn}
\end{equation}
Using equation (\ref{profilefence}) for $c(z)$, we obtain $S(n)\sim\lambda^{-4/3}n^{-1/2}$ for $z<\lambda$, and $S(n)\sim n^{-3/2}$ for $\lambda<z<\lambda_{2}$.
These two asymptotic expressions of $S(n)$ can be recast into a single formula:
\begin{equation}
S(n)\sim n^{-1/2}(g+n)^{-1}\label{Sn}
\end{equation}
where $g$, defined by equation (\ref{g}), is the mean number of monomers in a
strand of size $\lambda$. The
polyampholytes located at distance $z>\lambda_{2}$ are not adsorbed directly on the surface and should not be stretched in the AFM experiments; they thus do not participate to the loop size distribution. This loop size distribution \ref{Sn} is identical to the one derived by Dobrynin {\it et al} \cite{dob} when $n>g$; but for $n<g$, we find a power-law profile instead of a constant loop size, although we have to assume that all chains fold back for $z=\lambda$.

\section{Results and discussion}

\label{sec:res}
Images obtained by AFM on the
HS-(CH$_2$)$_2$-(CF$_2$)$_9$-CF$_3$ monolayer showed a remarkably ordered 2D
hexagonal lattice with a lattice constant equal to 0.59 nm on Au(111)/\-mica
\cite{tamada}; the surface density is then $3.3\,10^{18}$ thiols/m$^2$.
Keeping the same value for our systems and making the assumption that the ratio of charged thiols to (charged + neutral) thiols on 
the gold  surface remains equal to the ratio of charged thiols to (charged + neutral) thiols in the bulk solution used to process the gold 
surface (a very strong assumption indeed), we can estimate the surface charge density $\sigma_0$ 
for the various substrates. Using the estimated values of $\sigma_1$, $\sigma_2$ and $\sigma_3$ of 
Section \ref{sec:theory}, it appears that our experiments are performed in the {\em fence} regime. 
This is justified later on Figure \ref{estimation} and allows us to process our data according to the 
theoretical analysis exposed in \ref{sec:analyse}. The last assumption
supposes that there is no differential adsorption of the two thiols on gold.

\subsection{Loop size distribution on HS-(CH$_2$)$_2$-SO$_3^-$/HS-(CH$_2$)$_2$-CH$_3$}
\begin{figure}[ptb]
\centering{\resizebox{.451\textwidth}{!}{
  \includegraphics{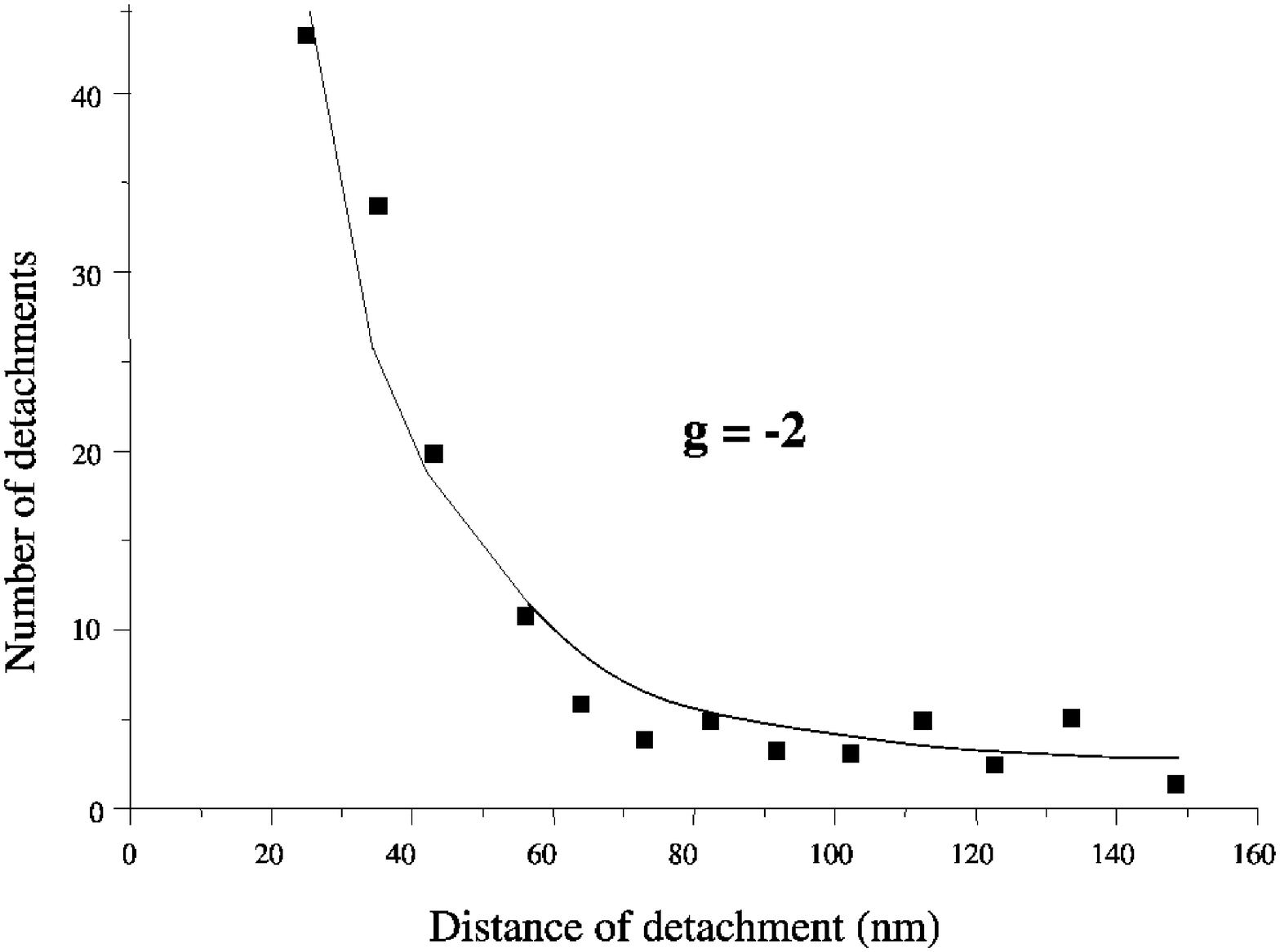}
}}
\caption{Loop size distribution on a charged surface obtained with HS-(CH$_2$)$_2$-SO$_3^-$ Na$^+$/HS-(CH$_2$)$_2$-CH$_3$ = 10 \% mol/mol. The full line represents the fit of the data using Eq. \ref{eqfit}.}
\label{tshort}
\end{figure}

The loop size distribution is shown on Figure \ref{tshort}. 

We find from the numerical fit that $g\simeq 0$ which means that no monomers are adsorbed in the layer of thickness $\lambda$. 
The loop size distribution $S(n)$ scales as n$^{-3/2}$ as in the studies of neutral polymers reported in \cite{senden,silber}. 
In fact, we have checked that the polyacrylamide homopolymer adsorbs 
on a surface covered by HS-(CH$_2$)$_2$-CH$_3$; short-chain thiols form less dense monolayers 
than long-chain thiols \cite{bain1} and are not efficient enough to prevent the polyampholyte adsorption; 
short-chain neutral thiols are thus not good candidates to investigate electrostatic effects.

\subsection{Loop size distribution on HS-(CH$_2$)$_2$-SO$_3^-$/HS-(CH$_2$)$_2$-(CF$_2$)$_7$-CF$_3$}
\begin{figure}[ht]
\centering{\resizebox{.45\textwidth}{!}{
  \includegraphics{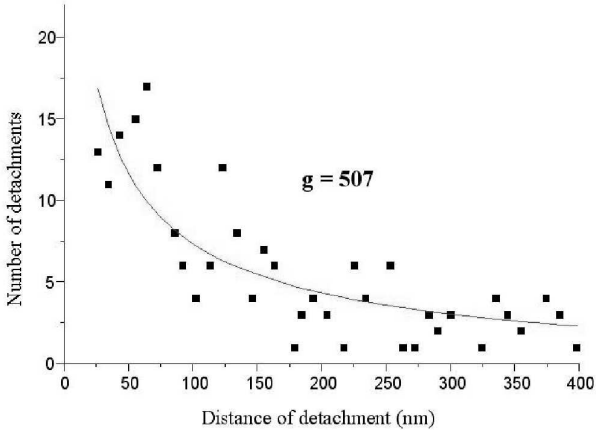}
}}
\caption{Loop size distribution on a charged surface prepared with HS-(CH$_2$)$_2$-SO$_3^-$ Na$^+$/HS-(CH$_2$)$_2$-(CF$_2$)$_7$-CF$_3$ = 0.1 \% mol/mol.  The full line represents the fit of the data using Eq. \ref{eqfit}.}
\label{t01}
\end{figure}

\begin{figure}[ht]
\centering{\resizebox{.45\textwidth}{!}{
  \includegraphics{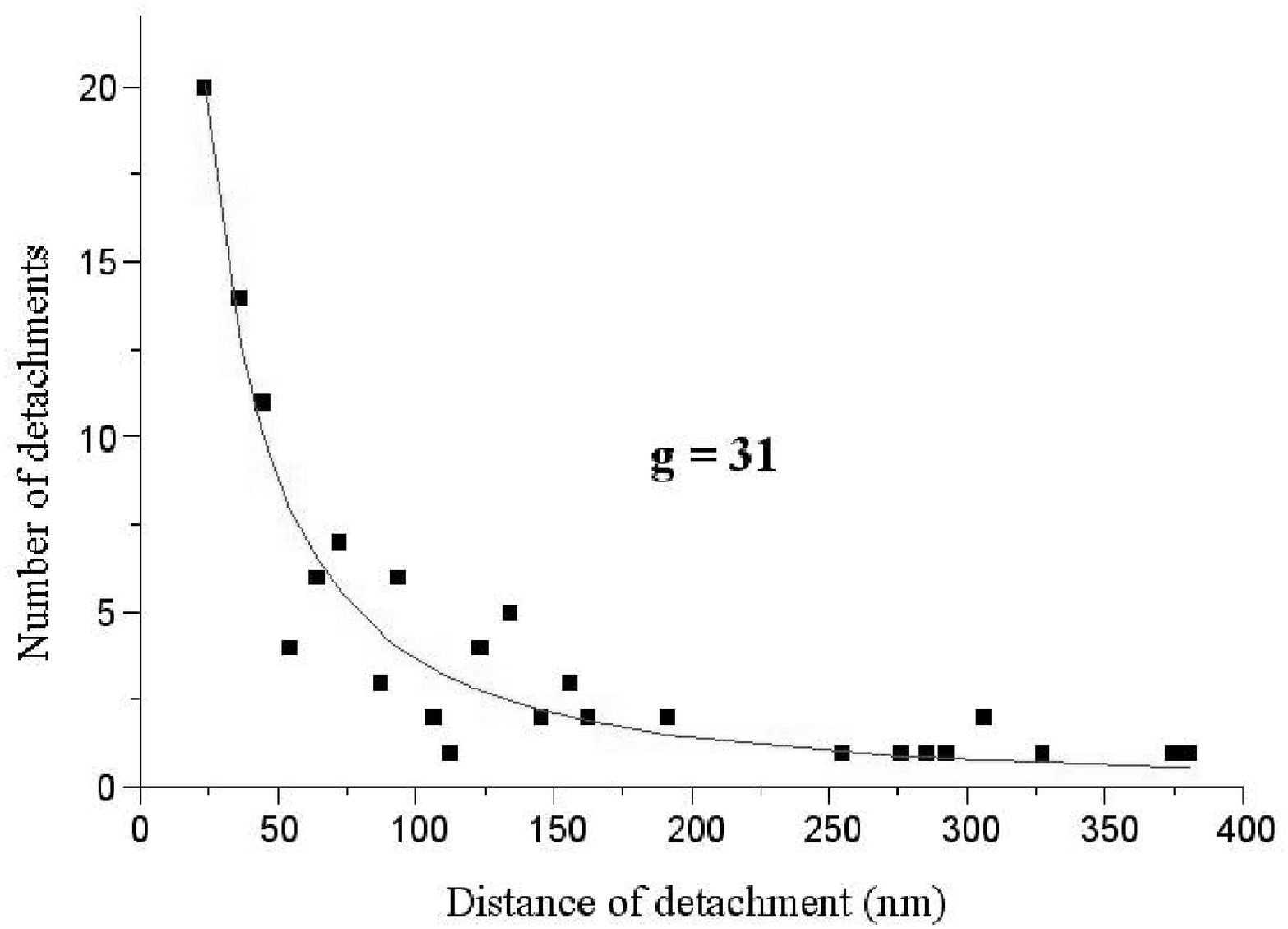}
}}
\caption{Loop size distribution on a charged surface prepared with HS-(CH$_2$)$_2$-SO$_3^-$ Na$^+$/HS-(CH$_2$)$_2$-(CF$_2$)$_7$-CF$_3$ = 3 \% mol/mol.  The full line represents the fit of the data using Eq. \ref{eqfit}.}
\label{t3}
\end{figure}

Giving up with HS-(CH$_2$)$_2$-CH$_3$, we turned to 
HS-(CH$_2$)$_2$-(CF$_2$)$_7$-CF$_3$ because of its longer chains and enhanced hydrophobicity. 
We have first checked that polyacrylamide does not
adsorb on the fluorinated thiols: no detachment peak has been observed on the force
curves. We cannot of course rule out from this observation that the polyampholyte cannot adsorb by non-electrotatic interactions between the fluorinated thiols and the chain skeleton; this possibility has not been investigated.  On the other hand, polyacrylamide adsorbs on a surface covered with HS-(CH$_2$)$_2$-SO$_3^-$, 
certainly via van der Waals interactions with the underlying gold surface, and consequently the polyampholyte
also adsorbs. In order to favour the electrostatic interactions  between the
charged monomers and the surface against
the hydrogen bonds between acrylamide and the charged thiols, weakly  charged surfaces have been prepared with millimolar solutions 
of thiols mixtures HS-(CH$_2$)$_2$-SO$_3^-$ Na$^+$/HS-(CH$_2$)$_2$-(CF$_2$)$_7$-CF$_3$ 
with molar ratios 0.1, 0.3, 1 and 3 \%. For the two extreme cases, the
loop size distribution and the fit given by equation (\ref{eqfit}) are shown on
Figures \ref{t01} and \ref{t3}. 

For each type of surface (0.1, 0.3, 1 and 3 \%), experiments have been repeated with four different substrates and the results 
of the fit (using Eq. \ref{eqfit}) are given in Table \ref{fittabl}.  The
different results obtained show that there is a charge
effect. We can deduce that the salt initially added to dissolve the polyampholyte chains has been effectively removed from the adsorbed
layer after rinsing, otherwise the field at the surface
would be screened after less than 1 nm (for an electrolyte concentration of 0.15 mol.l$^{-1}$) and then the
loop size distributions for different surfaces would be
identical. Moreover, the preferential adsorption of one thiol
in comparison with the other is limited, otherwise
the surfaces would be covered with the same type  of thiol and there
would not be any charge effect. This observation does not however allow to conclude that the rate of charged thiols at the
surface is the same as the one in volume.

\begin{table}
\begin{center}
\begin{tabular}[c]{|c|c|c|c|c|}
\hline

\% of charged thiol & 0.1 & 0.3 & 1 & 3\\
\hline
Expt. \#1 & 466 & 170 & 91 & 31\\
\hline
Expt. \#2 & 507 & 70 & 36 & 22\\
\hline
Expt. \#3 & 1051 & 302 & 132 & 96\\
\hline
Expt. \#4 & 666 & 395 & 22 & 4\\
\hline
$<g>$ & 672 & 234 & 70 & 38\\
\hline
$\sqrt{<g^2>-<g>^2}$ & 231 & 124 & 44 & 35\\
\hline
\end{tabular}
\vskip 5mm
\caption{Fitted values of $g$ for 4 different experiments with mean value $<g>$ and standard deviation $\sqrt{<g^2>-<g>^2}$.}
\label{fittabl}
\end{center}
\end{table}

\begin{figure}[ht]
\centering{\resizebox{.45\textwidth}{!}{
  \includegraphics{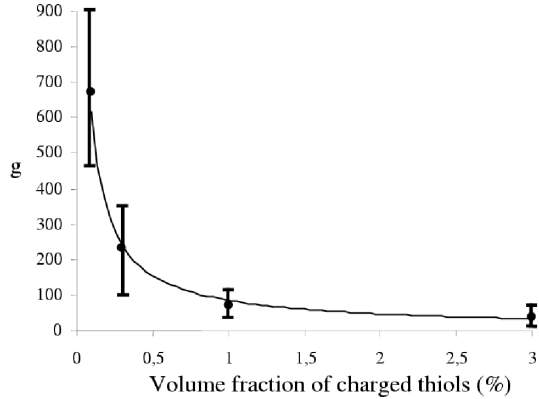}
}}
\caption{Mean number $<g>$ of monomers in a loop against the molar ratio of charged/(charged + neutral) thiols in volume. The full line is a power law fit (see text).}
\label{fit}
\end{figure}

The results for the mean number $<g>$ of monomers in a loop are also plotted on Figure \ref{fit} 
taking the standard deviation as experimental bars.

\begin{figure}[ptb]
\centering{\resizebox{.48\textwidth}{!}{
  \includegraphics{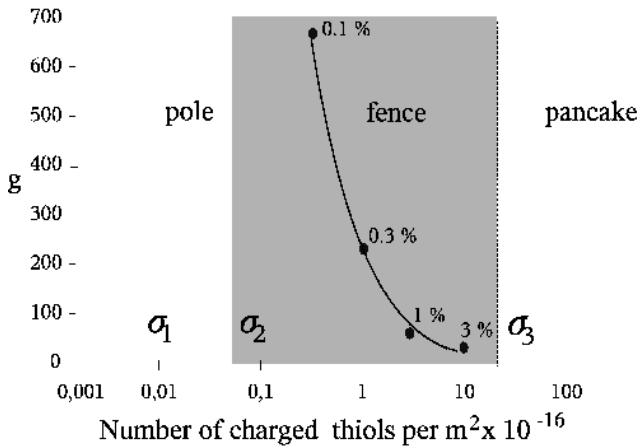}
}}
\caption{The various adsorption regimes of a polyampholyte on
a charged surface along with the number $g$ of monomers in a loop 
versus the volume fraction of charged thiols in the adsorption solution.}
\label{estimation}
\end{figure}

Figure \ref{estimation} shows that the AFM experiments
are in the fence regime; as predicted by theory, the number $g$ of monomers in a loop decreases with the surface charge.
$g$ scales like the ratio of charged/neutral thiols to the {-0.86} power; however, it seems difficult to compare it to the exponent
-1.33 given by Eq. \ref{g}, because the theoretical exponent links $g$ to the surface
charge and not the bulk fraction of charged thiol.

\subsection{Contact angle and electrokinetics measurements}
\begin{figure}[ht]
\centering{\resizebox{.45\textwidth}{!}{
  \includegraphics{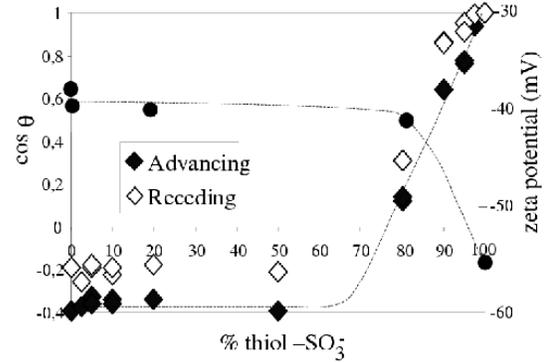}
}}
\caption{Water contact angles  ($\blacklozenge$,$\lozenge$) and  $\zeta$ potential ($\bullet$)  on HS-(CH$_2$)$_2$-SO$_3^-$/HS-(CH$_2$)$_2$-(CF$_2$)$_7$-CF$_3$ charged surfaces.}
\label{anglzeta}
\end{figure}

Contact angles and $\zeta$ potential for the charged surfaces  made with 
HS-(CH$_2$)$_2$-SO$_3^-$/HS-(CH$_2$)$_2$-(CF$_2$)$_7$-CF$_3$ are presented on Figure \ref{anglzeta}.

 The values
of $\zeta$ obtained for the monolayer of fluorinated thiols (-~37 and -~39 mV) are
close to the value measured by Chibowski and Waksmunki for polytetrafluoroethylene (CF$_{2}$)$_{n}$ \cite{teflon}.
 In bidistilled water, they
indeed found a potential equal to -~46.6 mV. The $\zeta$ potential of the monolayer of
charged thiols, -~55 mV, can be compared to that of a surface covered by
mercaptopropionic acid (HS-(CH$_2$)$_2$-COOH). By atomic force
measurements in a $10^{-3}$ mol.l$^{-1}$ KCl aqueous solution, Hu and Brad
\cite{bard} determined the potential $\psi_{d}$ above the counterion layer
of this surface; $\psi_{d}$ is indeed close to $\zeta$ \cite{hunter}. At  {pH $> 10$}, the acid is completely dissociated, and the potential
$\psi_{d}$ was -~62 mV, which is compatible with the streaming current measurement.

\begin{figure}[ht]
\centering{\resizebox{.45\textwidth}{!}{
  \includegraphics{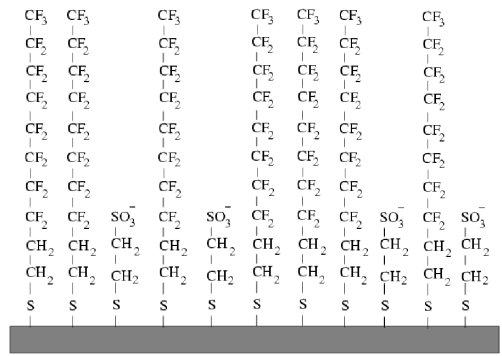}
}}
\caption{Schematic distribution of HS-(CH$_2$)$_2$-SO$_3^-$ Na$^+$ and HS-(CH$_2$)$_2$-(CF$_2$)$_7$-CF$_3$ on gold. }
\label{thiols}
\end{figure}

Figure \ref{anglzeta} shows that $\cos \theta$ and $\zeta$ are relatively well
correlated: their values are hardly modified for a bulk fraction of charged
thiol lower than 70-75 \%, and then strongly change. The
contact angle and the $\zeta$ potential are thus linked to the same phenomenon.
The plateau can be explained by i) a preferential adsorption of the fluorinated
thiol, ii) a phase separation between the two types of thiols, or iii) the conformations of the thiol alkyl chains. 
The first hypothesis is refuted by the AFM
experiments, which show a difference between the surfaces in the region of the
plateau (0.1 - 3 \%). 
We have not performed AFM imaging to investigate the second hypothesis and the possible presence of islands resulting from phase separation. Anyway the contact angle hysteresis
$\cos{\theta_{r}}-\cos\theta_{a}$ is limited and identical to the one of a pure
monolayer of fluorinated thiols, which is a very good indication of an homogeneous surface;  the $\zeta$ potential would also be extremely sensitive to the presence of charged islands. 
The third hypothesis seems to be the most likely. 
As shown by Figure \ref{thiols}, the HS-(CH$_2$)$_2$-SO$_3^-$ thiols
are much smaller than the fluorinated ones, and the charges that they carry are hidden
by the long neutral chains, which decreases their influence by a {\em screening} effect: the local reduced dielectric constant should also decrease the dissociation constant of the sulfonic groups of the smaller thiols, with the effect of a smaller effective smaller charge. 
This remains very  schematic, since thiolated chains can be tilted on the surface and the long
chains are unlikely to be in an all-trans conformation. However, AFM images of a monolayer of the fluorinated thiol have revealed 
extended conformations with a tilt angle of 22$^{\circ}$ \cite{naud}.

\begin{figure}[ht]
\centering{\resizebox{.45\textwidth}{!}{
  \includegraphics{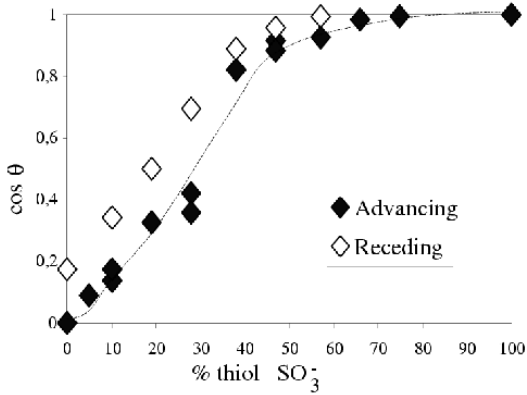}
}}
\caption{Contact angle of water on a HS-(CH$_2$)$_2$-SO$_3^-$/HS-(CH$_2$)$_2$-CH$_3$ monolayer (after \cite{annelaure}).}
\label{anglprop}
\end{figure}

Figure \ref{anglprop} represents data obtained by A.-L. Bernard, who
studied the surfaces HS-(CH$_2$)$_2$-SO$_3^-$/HS-(CH$_2$)$_2$-CH$_3$
\cite{annelaure}. Contrary to the former situation, the variation of
$\cos \theta$ with respect to the percentage of charged thiols in the bulk is
nearly linear, except for highly charged surfaces. The two thiolated molecules have the same length and there is no screening
effect.

\subsection{Interpretation}
\begin{figure}[ht]
\centering{\resizebox{.45\textwidth}{!}{
  \includegraphics{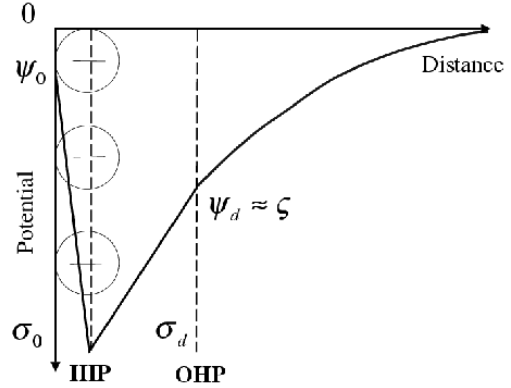}
}}
\caption{Preferential adsorption of anions on a weakly charged surface; electrostatic potential profile in solution. 
The inner Helmholtz plane (IHP) is defined by the plane of adsorption of desolvated ions and the outer Helmholtz plane (OHP)
by the plane for which the solvated ions are closest to the surface.}
\label{adprefa}
\end{figure}

A striking feature of  Figure \ref{anglzeta} is the
non-zero value of the $\zeta$ potential of the monolayer of fluorinated thiols, which
are neutral molecules. Similar streaming current measurements on a monolayer
of octadecanethiol \cite{zetathiol} have showed that the $\zeta$ potential
varies with pH and salt concentration. The
proposed explanation is a preferential adsorption of the hydroxide ions on
the surface. In the case of the fluorinated monolayer, a preferential adsorption
of anions certainly takes place, but it is difficult to decide whether the
adsorbing ions are OH$^{-}$ or Cl$^{-}$. The phenomenon is sketched on
Figure \ref{adprefa}. The $\zeta$ potential is approximately equal to the potential
$\psi_{d}$\ at the outer Helmoltz plane (OHP) and thus non-negligible due to the adsorption of anions, whereas the surface potential
$\psi_{0}$ remains weak \cite{hunter}.

\begin{figure}[ptb]
\centering{\resizebox{.45\textwidth}{!}{
  \includegraphics{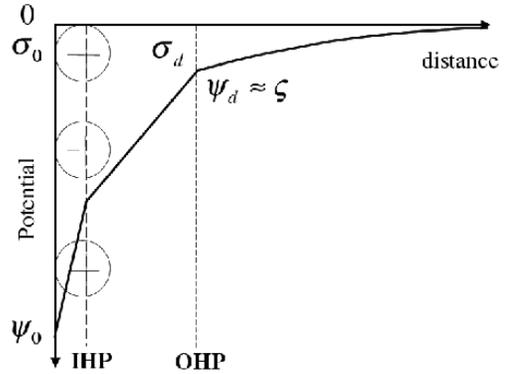}
}}
\caption{Screening by counterions of a highly charged surface; electrostatic potential profile in solution.}
\label{adprefc}
\end{figure}

The surface charge $\sigma_{d}$\ of the OHP plane close to the monolayer of
charged thiols can be obtained with the hypothesis $\zeta\simeq\psi_{d}$ and
the relation given by the Gouy-Chapman-Stern-Graham theory: 
$\sigma_{d} = 4ce\kappa^{-1}\sinh(e\psi_{d}/2k_BT)$, where $c$ is the salt concentration, and
$\kappa^{-1}$ the Debye length \cite{hunter}.  In our streaming current
measurements on a monolayer
of charged thiols, c = $10^{-3}$ mol.l$^{-1}$, $\kappa^{-1}\simeq$ 10 nm,
$\psi_{d}$ = -~55 mV. The obtained value $\sigma_{d}$
$\simeq-4.9\,10^{-3}$ C.m$^{-2}$ corresponds to a charge density equal to
$3.1\,10^{16}$ negative charges / m$^{2}$, much lower than the
surface density expected for a monolayer of charged thiols: low energy
helium diffraction on a
HS-(CH$_2$)$_9$-CH$_3$ monolayer \cite{helium} and surface acoustic wave
studies on a HS-(CH$_2$)$_6$-CH$_3$ monolayer
\cite{acoustic} both established that the surface density of the molecules
of the  thiolated monolayer is around $5\,10^{18}$
thiols/m$^{2}$.  If we suppose that the density is the same for the charged
thiol monolayer, the large difference between this
value and the former charge density implies a strong screening by the counterions (cations) in
solution, represented on Figure \ref{adprefc}. A similar effect was also
inferred by Hu and Hard after AFM experiments on thiol monolayers \cite{hu}.
We would like to stress here that it is  anyway difficult to determine the actual charge that rules the electric potential of a charged surface. 
The determination of the charge is in fact model-dependent; the actual charge density is likely to be between the values estimated from 
direct (crystallographic) measurements (such as AFM \cite{tamada}) and electrokinetic
measurements.

\begin{figure}[ht]
\centering{\resizebox{.45\textwidth}{!}{
  \includegraphics{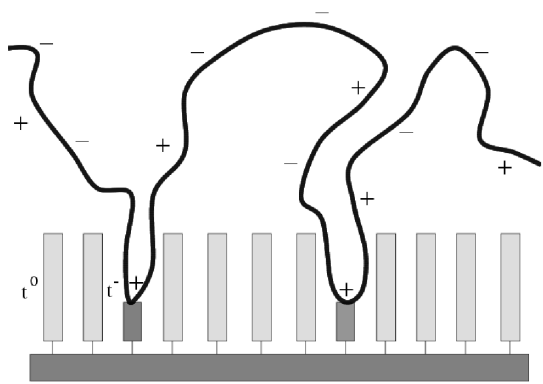}
}}
\caption{Schematic sketch of the adsorption of the polyampholyte on a 
HS-(CH$_2$)$_2$-SO$_3^-$ Na$^+$ / HS-(CH$_2$)$_2$-(CF$_2$)$_7$-CF$_3$ monolayer.}
\label{polyamph}
\end{figure}

Concerning the surfaces used in AFM experiments (0.1, 0.3, 1 and 3 \% of
charged thiols in the bulk), contact angle and electrokinetics measurements 
do not show any difference with respect to a monolayer of
100 \% neutral fluorinated thiols (see Figure \ref{anglzeta}): $\theta \simeq 115^{\circ}$ 
and $\zeta\simeq - 38$ mV. The study of the detachment distances in AFM experiments
have however revealed a charge effect in qualitative agreement with the theory.
These apparently contradictory observations can be explained by making the
following assumption: the few short-chain charged thiols are hidden by the long
fluorinated ones. Contact angle and electrokinetics measurements give then the
feeling that these surfaces are covered by a monolayer of neutral thiols.
These measurements are indeed realized at a mesoscopic scale, whereas the AFM
ones are realized at a microscopic scale. Even if the contact angle and the
zeta potential \textit{see} only the neutral fluorinated thiols, polyampholytes have
the ability to find their way between them to adsorb on the charged
thiols (Figure \ref{polyamph}).

\section{Conclusion}
\label{sec:concl}
The adsorption of a polyampholyte has been studied on various HS-(CH$_2$)$_2$-SO$_3^-$/HS-(CH$_2$)$_2$-(CF$_2$)$_7$-CF$_3$ monolayers by 
AFM. The statistical analysis of the loop size
distribution in the adsorbed layer reveals that the number of monomers
in a loop decreases, in qualitative agreement with theoretical predictions
\cite{dobrynin,obukhov}. In order to measure  the charge density $\sigma_0$\ of the
surfaces, we have performed streaming current measurements, which gave us the
$\zeta$ potential of the surfaces. Because of a phenomenon of preferential
adsorption of some ions, the value of $\sigma_{0}$ could not be deduced from
$\zeta$. For instance, the existence of a non-zero $\zeta$ potential on the
monolayer of fluorinated thiols is difficult to interpret; an explanation could be
that the bond between the fluorocarboned part of the thiol and its
hydrocarboned part creates a strong dipole \cite{dipole}; this could attract
anions of the electrolyte by induced-type interactions. The dipoles oriented
in the bonds -CH$_{2}$-CF$_{2}$- moreover constitute a layer of dipoles which
could contribute to this attraction.

The contact angle and $\zeta$ potential measurements are well correlated, and give
the same value for the monolayer of fluorinated thiols than for weakly charged
surfaces. AFM experiments on these surfaces however show different
structures of the polyampholyte adsorbed layer. This apparent paradox is
explained  by the following interpretation: the short-chain charged thiols are hidden by
the long-chain fluorinated ones, but the polyampholyte has the ability to find its way
between them to adsorb. This hypothesis could be checked by infrared or Raman
surface spectrophotometry and, if confirmed, could lead to interesting
applications, such as the recognition by charged polymers of charged groups which
are undetectable at a mesoscopic scale by a simple wetting or
electrokinetics study.

\section{Acknowledgments}
The authors are thankful to Drs. Fran\c{c}oise Candau,  Franck
Cl{\'e}ment, Georges Debr\'egeas, Albert
Johner, Marie-Pierre Kraft, Hans Lyklema, Philippe Mesini, Joseph Selb and Tim Senden for fruitful discussions.

\bibliographystyle{unsrt}
\bibliography{polyamp}

\end{document}